\documentclass[a4paper,fleqn,usenatbib]{mnras}

\usepackage{newtxtext,newtxmath}

\usepackage[T1]{fontenc}
\usepackage{ae,aecompl}

\usepackage{graphicx}	
\usepackage{amsmath}	
\usepackage{amssymb}	

\title[Eclipse mapping of SDSS\,J0926+3624]{Mapping the accretion disc of the short period eclipsing binary SDSS\,J0926+3624}

\author[Schlindwein \& Baptista]{
Wagner Schlindwein$^{1}$
\& Raymundo Baptista$^{1}$
\\
$^{1}$Departamento de F\'{i}sica, Universidade Federal de Santa Catarina, Campus Trindade, 88040-900 Florian\'{o}polis, SC, Brazil
}

\date{Accepted XXX. Received YYY; in original form ZZZ}

\pubyear{2018}

\begin{document}
\label{firstpage}
\pagerange{\pageref{firstpage}--\pageref{lastpage}}
\maketitle

\begin{abstract}
We report the analysis of time-series of optical photometry of SDSS\,J0926+3624 collected with the Liverpool Robotic Telescope between 2012 February and March while the object was in quiescence. We combined our median eclipse timing with those in the literature to revise the ephemeris and confirm that the binary period is increasing at a rate $\dot{P}=(3.2 \pm 0.4)\times 10^{-13} \, s/s$. The light curves show no evidence of either the orbital hump produced by a bright spot at disc rim or of superhumps; the average out-of-eclipse brightness level is consistently lower than previously reported. The eclipse map from the average light curve shows a hot white dwarf surrounded by a faint, cool accretion disc plus enhanced emission along the gas stream trajectory beyond the impact point at the outer disc rim, suggesting the occurrence of gas stream overflow/penetration at that epoch. We estimate a disc mass input rate of $\dot{M}=(9 \pm 1)\times 10^{-12}\,M_\odot \,yr^{-1}$, more than an order of  magnitude lower than that expected from binary evolution with conservative mass transfer.
\end{abstract}




\section{Introduction}

AM Canum Venaticorum (AM~CVn) stars are ultracompact binaries with orbital periods ranging from 5.4 to 65\,min and optical spectra dominated by helium lines \citep[see e.g.,][]{Nelemans2005,Ramsay2007,Roelofs2010}. They consist of a white dwarf (the primary) accreting matter via an accretion disc from a significantly less massive, hydrogen-deficient donor star (the secondary). In order to fills its Roche lobe and sustain the mass transfer process, the donor star needs to be at least partially degenerate. AM CVn stars offer new insights into the formation and evolution of binary star systems, with the short periods implying at least one common envelope phase in the history of the binary, and the chemical composition suggesting helium white dwarfs or cataclysmic variables (CVs) with evolved secondaries as possible progenitors (\citealp[see also]{Nelemans2001,Nelemans2010} \citealp{Marsh2011,Kulkarni2010} for the recent identification of a possible AM CVn progenitor).

SDSS\,J0926+3624 (hereafter J0926) was the first eclipsing AM~CVn star and is also one of the shortest period eclipsing binary known \citep{Anderson2005}. Its light curve displays deep ($\sim 2$~mag) eclipses every 28.3~min, which lasts for $\sim 2$~min, as well as $\sim 2$~mag amplitude outbursts recurring on timescales of $\sim 100 - 200$~d \citep[][hereafter C11]{Copperwheat2011}. Superhumps were seen in its light curves several days after the end of an outburst \citep[C11;][hereafter Z14]{Szypryt2014}; these are believed to result from the tidal interaction between the mass-donor star and an elliptical precessing disc \citep[e.g.,][]{Whitehurst1988,Hirose1990}. C11 modeled the eclipse light curves to estimate masses and radii of both stars, a mass ratio $q= M_2/M_1= 0.041\pm 0.002$ and an inclination $i= 82\fdg 6 \pm 0\fdg 3$, a white dwarf temperature of $T_{wd}= 17000\,K$ and a corresponding distance estimate of $(465\pm 5)$\,pc.  Z14 found that the orbital period of J0926 is increasing at a rate $\dot{P}= (3.07 \pm 0.56) \times 10^{-13} \, s/s$ and, as a consequence, inferred a conservative mass transfer rate $\dot{M} \simeq 1.8 \times 10^{-10}\, M_\odot\, yr^{-1}$. This is in agreement with the $\dot{M}= (1.4 \pm 0.3) \times 10^{-10}\, M_\odot \, yr^{-1}$ predicted by \citet{Deloye2007} based on their stellar evolutionary calculations (corrected for the $q=0.041$ value of C11).

Here we report additional high-speed optical photometry of J0926, used to confirm its orbital period change and to disentangle its optical light sources with the aid of eclipse mapping techniques. The observations are described in Sect.~\ref{observstions}; data analysis and results are reported in Sect.~\ref{analysis} and summarized in Sect.~\ref{summary}.


\section{Observations and data reduction}\label{observstions}

Time series of high-speed CCD photometry of J0926 in the $V+R$ passband were obtained with the RISE camera ($512 \times 512$ pixels, $1."07$\;pixel$^{-1}$ in the $2\times 2$ binning mode) attached to the 2.0\,m Liverpool Robotic Telescope (LRT), in the Canary Islands, during 2012 February-March while the object was in quiescence and near the end of a 4.6\,yr long period without recorded outbursts. The CCD camera is operated in frame-transfer mode, with negligible dead time between exposures.

The observations are summarized in Table~\ref{tab_journal}. The fourth column lists the range of eclipse number (E), the fifth column lists the seeing range, and the last column gives an estimate of the quality of the observations. Each run consists of a 2\,h long sequence of 442 exposures at a time resolution of 16\,s covering 4 orbits of the binary; the data set comprises a total of 20 eclipse light curves. All light curves were obtained with the same set of instrument and telescope, which ensures a high degree of uniformity to the data set. Poor weather and seeing conditions forced us to discard about half of the data on Mar~21.

\begin{table}
 \begin{minipage}{84mm}
 	\centering
	\caption{Journal of Observations}
	\label{tab_journal}
	\begin{tabular}{cccccc}
		\hline
		Date & \multicolumn{2}{|c|}{UT} & E \footnote{With respect to the ephemeris of eq.~(\ref{eq_ephemeris}).} & Seeing & Sky \footnote{Sky: (A) Photometric; (B) good; (C) poor (large transparency variations, thin clouds,or both).} \\
		(2012) & Start & End & (cycle) & (arcsec) &  \\
		\hline
		Feb 23 & 22:15 & 00:13 & 111131 - 111135 & 1.0 - 1.9 & A \\
		Mar 18 & 22:49 & 00:47 & 112353 - 112357 & 1.3 - 2.5 & B \\
		Mar 21 & 21:36 & 23:34 & 112503 - 112507 & 1.2 - 3.0 & C \\
		Mar 23 & 21:17 & 23:15 & 112604 - 112608 & 1.1 - 1.8 & A \\
		Mar 25 & 22:41 & 00:38 & 112709 - 112713 & 1.1 - 1.7 & A \\		
		\hline
	\end{tabular}
 \end{minipage}
\end{table}

As part of the observatory pipeline, data from the LRT are distributed to the user already corrected from the instrumental effects of dark current, bias and flat-field. Aperture photometry was used to extract fluxes for a reference star, the variable, and for a set of nearby comparison stars with scripts using the apphot/IRAF\footnote{IRAF is distributed by the National Optical Astronomy Observatories, which are operated by the Association of Universities for Research in Astronomy, Inc., under cooperative agreement with the National Science Foundation.} routines in order to derive light curves of differential photometry. In order to flux calibrate the light curves, we used the aperture photometry of the comparison field star LP 260-21, which has calibrated fluxes in the $B,\ V,\ R,\ J,\ H$ and $K_s$ passbands in the SIMBAD database \citep{Wenger2000}. We then performed synthetic photometry with the synphot/IRAF package to find the best-fit match of the Bruzual-Persson-Gunn-Stryker atlas of spectrophotometric stars \citep{Strecker1979,Gunn1983} to the SIMBAD observed fluxes of LP\,260-21, and convolved the best-fit M4\,V spectrum with the response of the RISE camera $V+R$ passband to find a calibrated flux of $0.78\pm 0.08$~mJy for this star. The calibrated flux of the reference star was then used to transform the light curves of the variable and of the comparison stars from magnitude difference to absolute flux. This absolute flux calibration has an estimated uncertainty of about 10 per cent. On the other hand, analysis of the relative flux of the comparison stars in all data sets indicates that the internal error of the photometry is less than 3 per cent. The error in the photometry of the variable is derived from the photon-count noise and is transformed to flux units using the same relation applied to the data. The individual light curves have typical signal-to-noise ratios of $S/N\simeq 10$. Exposure times were transformed from Universal Time (UTC) to Terrestrial Dynamical Time (TDT) and to the Baricentric reference frame (BJDD), according to the code of \citet{Eastman2010}, to account for the light travel time effect.


\section{Data analysis and results}\label{analysis}

\subsection{Light curves and revised ephemeris}

The individual light curves were phase-folded according to the linear ephemeris of C11,

\begin{equation}
 T_\mathrm{mid}(BJDD) = 2\,453\,796.445\,5191(5) +  0.019\,661\,272\,89(2)\times E \,\, ,
 \label{eq_ephemeris}
\end{equation}

\noindent where $T_\mathrm{mid}$ is the primary mid-eclipse time and $E$ is the binary cycle. At first, we grouped the data sets to obtain separate average light curves for each of the observing nights. However, since the average light curves of all nights show the same morphology, brightness level, eclipse shape and depth, we combined all data to obtain a single average light curve with increased S/N.

We used a phase-folded, concatenated light curve to obtain a single mid-eclipse time from the whole data set. The concatenated light curve was median filtered with a boxcar of phase width 0.015 and fitted with a model eclipse light curve of the white dwarf (see Sect.~\ref{eclipsemapping}) to find a white dwarf (WD) mid-eclipse phase of $\phi_0= +0.0009 \pm 0.0005$. The measured eclipse time is delayed by 1.5\,s with respect to Eq.~\ref{eq_ephemeris}, in very good agreement with the results of Z14. We adopted the median eclipse cycle of our median observing date, used Eq.~\ref{eq_ephemeris} to compute its predicted mid-eclipse time and added the measured $\phi_0$ value to obtain an equivalent observed mid-eclipse timing of $T_\mathrm{mid}(E=112\,505)= \mathrm{BJDD} \; 2\,456\,008.437\,044(10)$.

We evaluated the influence of the relatively long 16\,s exposure time of the individual light curves (with respect to the $\sim 120$\,s total width of the eclipse) on our ability (i) to precisely measure mid-eclipse timings and (ii) to recover the eclipse shape (for the purpose of eclipse mapping) with the following experiment. We simulated the eclipse of the surface brightness distribution in the upper panel of Fig.~\ref{fig_mapeamento} at a time resolution of 0.5\,s to construct a base light curve, to which a phase offset of $\phi_0= +0.0009$ was added. We then created two sets of 20 light curves from the base light curve (by binning the data at a time resolution of 16\,s with random starting times), one noise-free data set and one data set with 10 per cent gaussian noise added to each data point ($S/N=10$). We repeated the above procedure to estimate a mid-eclipse timing from the noisy, simulated data set and were able to recover the added phase offset with the same precision of the real data, namely, $\phi_0= +0.0009 \pm 0.0005$.  Furthermore, the comparison of the noise-free data set with the base light curve indicates that the long exposure time of the observations lead to systematic, small differences in eclipse shape at the 1 per cent level, which are negligible in comparison with the 10 per cent noise affecting the real data. This shows that the uncertainties in the mid-eclipse timing and in the eclipse shape are largely dominated by the noise in the light curves, with negligible contribution from the relatively long exposure time of the observations.

In order to derive a revised ephemeris for J0926, we combined our timing with those in the literature. We measured the Observed-minus-Calculated (O-C) values of C11 from their Fig.\,12; we computed median (O-C) values for each of their 6 nights of observations and used their ephemeris with a representative cycle for each night to derive equivalent BJDD eclipse times. The same procedure was used to derive an equivalent, median eclipse time from Fig.\,4 of Z14. We excluded the timings of their first two nights of observations from this analysis as the object was in outburst and the eclipses were much shallower than usual, being systematically later than those of the last night. \footnote{Possibly the consequence of the significant out-of-eclipse modulation reminiscent of superhumps, particularly in the 2012 Dec 8 data.} The results are listed in Table\,\ref{timings}. Table~\ref{efemerides} presents the parameters of the best-fit linear and quadratic ephemerides to these timings with their 1-$\sigma$ formal errors quoted, together with the root-mean-square value of the residuals, $\sigma$, and the reduced $\chi^2_\nu$, where $\nu$ is the number of degrees of freedom. These fits assume equal errors of $2.2\times 10^{-6}\,d$ to the data points, which ensures a unity $\chi^2_\nu$ for the quadratic ephemeris.

\begin{table}
\centering
 \caption{Median mid-eclipse times of J0926.}
 \label{timings}
  \begin{tabular}{rccc}
  \hline
  E & BJDD & (O-C) & Reference \\ [-0.5ex]
  (cycle) & (2450000+) & (s) \\
  \hline
      7 & 3796.583142 & $-0.10$ & C11 \\
     55 & 3797.526887 & $+0.24$ & C11 \\
    105 & 3798.509946 & $-0.15$ & C11 \\
  52750 & 4833.577663 & $+0.22$ & C11 \\
  52803 & 4834.619706 & $-0.15$ & C11 \\
  52857 & 4835.681416 & $-0.05$ & C11 \\
 112505 & 6008.437044 & $-0.05$ & This work \\
 125963 & 6273.038464 & $+0.05$ & Z14 \\
  \hline
  \end{tabular}
\end{table}

\begin{table}
\centering
 \caption{Ephemerides for mid-eclipse times of J0926.}
 \label{efemerides}
  \begin{tabular}{ll}
  \hline
 {\bf Linear ephemeris:} \\ [-0.5ex]
 $\mathrm{BJDD} = T_0 + P_0 \, E$ \\
 $T_0= 2\,453\,796.445\,510(\pm 1)\,d$ & $P_0= 0.019\,661\,273\,12(\pm 2)\,d$ \\
 $\chi^2_{\nu_1}= 9.56 \, , \nu_1= 6$  & $\sigma_1= 6.23 \times 10^{-6}\,d$ \\
 [1ex]
 {\bf Quadratic ephemeris:} \\ [-0.5ex]
 $\mathrm{BJDD} = T_0 + P_0 \, E + c \, E^2$ \\
 $T_0= 2\,453\,796.445\,514(\pm 1)\,d$ & $P_0= 0.019\,661\,272\,75(\pm 5)\,d$ \\
 $c= (+3.18 \pm 0.44)\times 10^{-15}\,d$ & $\sigma_2= 1.85 \times 10^{-6}\,d$ \\
 $\chi^2_{\nu_2}= 1.00 \, , \nu_2= 5$ \\
  \hline
  \end{tabular}
\end{table}

\begin{figure}
	\includegraphics[width=0.36\textwidth,angle=270]{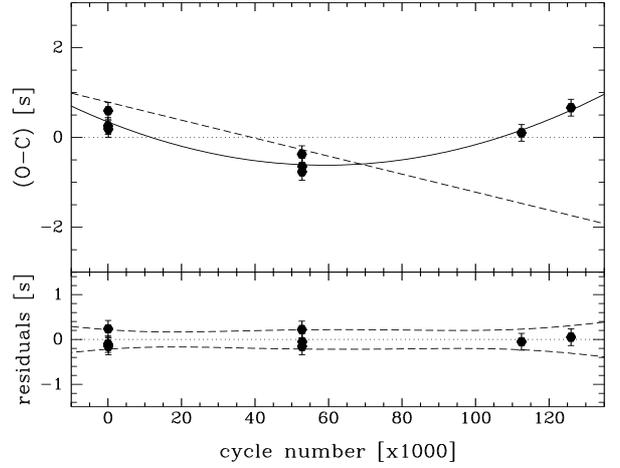}
    \caption{\textit{Top:} (O-C) diagram of J0926 with respect to the best-fit linear ephemeris of Table~\ref{efemerides}. The dashed line depicts the linear ephemeris of C11 and the solid curve shows the best-fit quadratic ephemeris. \textit{Bottom:} The residuals with respect to the best-fit quadratic ephemeris of Table~\ref{efemerides}. The dashed lines indicate the 2-$\sigma$ limits on the ephemeris, with the covariances taken into account.}
    \label{fig_oc}
\end{figure}

The top panel of Fig.~\ref{fig_oc} shows the (O-C) diagram with respect to the best-fit linear ephemeris and the lower panel shows the residuals with respect to the best-fit quadratic ephemeris. The quadratic fit provides a much better description of the data than the linear fit, underscoring the findings of Z14. The significance of adding an additional term to the linear ephemeris was estimated with the F-test \citep{Pringle1975}. The quadratic ephemeris has a statistical significance of 99.9 per cent with $F(1,6)=51.7$. The orbital period of J0926 seems to be increasing at a rate $\dot{P}= (3.24 \pm 0.43)\times 10^{-13} \, s/s$. The inferred $\dot{P}$ value is consistent with that found by Z14 within the uncertainties. The observed period increase could possibly be the consequence of binary evolution through angular momentum loss caused by gravitational radiation (Z14). However, given that every eclipsing CV with a well-sampled (O-C) diagram covering at least a decade of observations shows cyclical period changes \citep[e.g.,][]{Borges2008}, the observed period increase could alternatively be part of a cyclical period modulation of amplitude several seconds on a decade long timescale. Given the precision with which eclipses can be measured in this binary, a further decade of (regular) observations will suffice to clarify these possibilities.

\begin{figure}
	\includegraphics[width=\columnwidth]{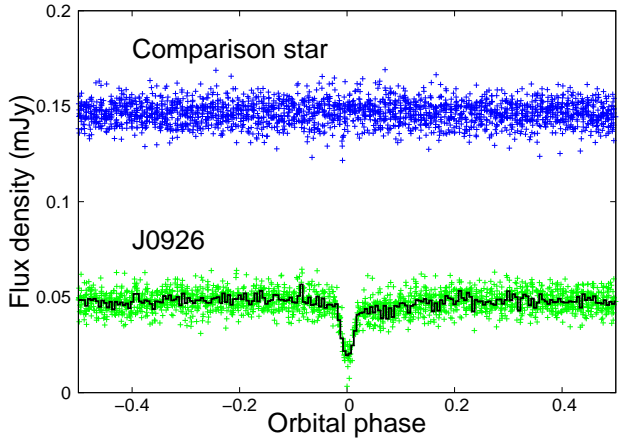}
    \caption{Phase-folded light curves of J0926 (green crosses) and of a nearby comparison star (blue crosses). The solid line is the median light curve at a phase resolution of 0.005 cycle.}
    \label{fig_light_curves}
\end{figure}

\begin{figure}
	\includegraphics[width=\columnwidth]{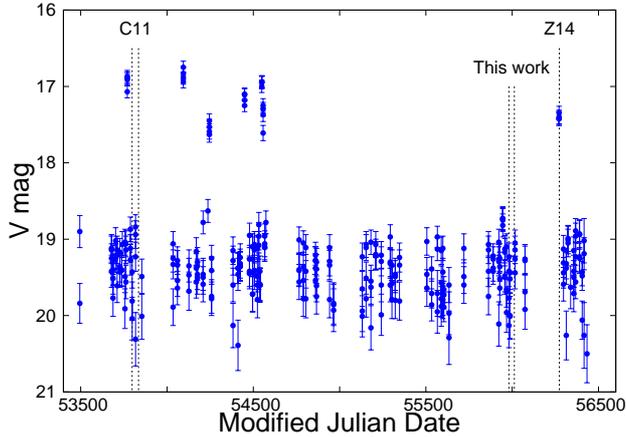}
    \caption{J0926 observations made by the CRTS for the interval from 2005 January 29 to 2013 November 3. Vertical dotted lines mark the epoch of the observations of C11, Z14 and of this work. Our observations frame the 4.6 yr long interval without recorded outbursts.}
    \label{fig_crts}
\end{figure}

Flux calibrated and phase-folded light curves of J0926 and of a comparison star are shown in Fig.~\ref{fig_light_curves}. An offset of $\phi_0= -0.0009$ was applied to the data to make the WD eclipse center coincide with phase zero. The average light curve shows the deep, narrow and partial eclipse of the WD at disc center superimposed on a broad, shallow eclipse of the disc plus incoming gas stream displaced towards positive phases. There is no evidence of superhump or of an orbital hump produced by anisotropic emission from a bright spot at disc rim. The average out-of-eclipse flux level is $0.053\pm 0.005$~mJy, $\sim 20$ per cent lower than that obtained by C11 after subtracting the superhump contribution from their light curves. The lack of evidence of superhumps and the clear differences in the light curve morphology with respect to the data of C11 and Z14 is not surprising. According to the Catalina Real-Time Transient Survey data \citep[CRTS,][]{Drake2009}, J0926 had no recorded outburst during the $\sim 4.6$\,yr period between  MJD 54550 and 56250 (Fig.~\ref{fig_crts}). Our data set frames this extended quiescent state. The last recorded outburst occurred $\sim 1450\,d$ ($\sim 4\, yr$) before our observations. Both the 2006 and the 2009 observations of C11 occurred at an epoch where the binary showed frequent outbursts, and the observations of Z14 occured right after the first recorded outburst at the end of this long quiescent state.

We end this section noting that the small inferred $\dot{P}$ value inplies a change in eclipse center of about $1 \, \mu s$ over the time span ($\sim 30$\,d) of our observations, several orders of magnitude lower than the uncertainty in the measured median eclipse timing.


\subsection{Search for periodicities}

We searched for periodicities in the light curves by computing a Lomb-Scargle periodogram \citep{Press1992} for each separate night. We subtracted the combined, average light curve of all nights from each individual light curve to eliminate the DC component, and we masked data in the phase range $(-0.1,+0.1)$ to avoid the eclipse. Amplitude levels in the power spectrum were evaluated by injecting artificial signals of known amplitude in the time series. The periodograms show no prominent signal at orbital $(f_\mathrm{orb}=
50.9\,d^{-1})$, superhump ($f_\mathrm{sh}=50.4\,d^{-1}$, C11), or at any other frequency in the range $f< 2000\,d^{-1}$. We find upper limits of 2 and 3 percent for the semi-amplitude of any superhump or orbital hump periodicity in the light curves, respectively for the nights of Feb 23, Mar 23-25 and the nights of Feb 18-21 (the lower quality nights), in agreement with the observed lower brightness level and the lack of evidence for out-of-eclipse modulation in the light curves (Fig.~\ref{fig_light_curves}).

\subsection{Eclipse mapping}\label{eclipsemapping}

We applied eclipse mapping techniques \citep{Horne1985,Baptista2016} to the average light curve of J0926 to solve for a map of the surface brightness distribution of its accretion disc and for the flux of an additional uneclipsed component.

Our eclipse map is a flat Cartesian grid of $51 \times 51$ pixels centered on the primary star with side $2R_{L_1}$ (where $R_{L_1}$ is the distance from the disc center to the inner Lagrangian point). The eclipse geometry is defined by the mass ratio $q$ and the inclination $i$. We adopted the parameters of C11, $R_{L_1} = (0.231 \pm 0.009)\,R_\odot$, $q = 0.041 \pm 0.002$ and $i=82.6^\circ \pm 0.03^\circ$, which correspond to a WD eclipse phase width of $\Delta\phi=0.022$. This combination of parameters ensures that the WD is at the center of the map.

Our eclipse mapping code implements the scheme of double default functions, $D_+\,D_-$, simultaneously steering the solution towards the most nearly axi-symmetric map consistent with the data ($D_+$), and away from the criss-crossed arcs along the edges of the shadow of the occulting, mass-donor star ($D_-$) \citep{Spruit1994,Baptista2005,Baptista2016}. It is optimized to recover asymmetric structures in eclipse maps such as spiral arms and enhanced gas stream emission. The positive default function is a polar Gaussian with radial and azimuthal blur widths of $\Delta r= 0.02\,R_{L_1}$ and $\Delta \theta = 30^\circ$, respectively. The negative default function is a Gaussian along the ingress/egress arcs of phase width $\Delta \phi = 0.01$.

\begin{figure}
	\includegraphics[width=\columnwidth]{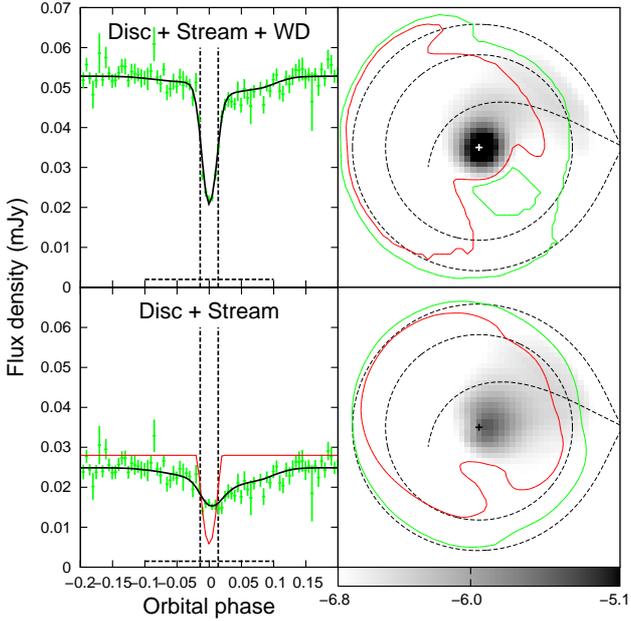}
    \caption{\textit{Upper left panel:} Data (green crosses with error bars) and model (black line) light curves. \textit{Lower left panel:} WD-subtracted data (green crosses with error bars) and model (black line) light curves. The red solid line depicts the model WD eclipse light curve. Horizontal dashed lines indicate the uneclipsed flux in each panel. Vertical dashed lines mark the ingress/egress phases of the WD. \textit{Upper right panel:} Surface brightness distribution of the combined disc+stream+WD map in a logarithmic grayscale. \textit{Lower right panel:} WD-subtracted surface brightness distribution. Regions inside the green/red contour lines are above the 2- and 3-$\sigma$ confidence levels, respectively. Dashed lines depict the primary Roche lobe, a disc of radius $0.65\,R_{L_1}$, and the ballistic stream trajectory. The horizontal bar shows the logarithmic intensity grayscale; brighter regions are darker.}
    \label{fig_mapeamento}
\end{figure}

The input, average light curve was obtained by slicing the combined data in phase bins of width 0.005 and by computing the median flux for each bin. The median of the absolute deviations with respect to the median flux was taken as the corresponding uncertainty at each bin (Fig.~\ref{fig_mapeamento}, upper panel). We also used the eclipse geometry to model the contribution of the WD to the light curve, assuming a DB white dwarf with $M_{WD}= 0.85\,M_\odot$, $\log g = 8.33$, $T_{WD}= 17000\,K$ and a distance of 465\,pc (C11), taking into account the effects of limb darkening \citep{Gianninas2013}. The WD model light curve was subtracted from the average light curve and the resulting curve was input to the eclipse mapping code in order to isolate the surface brightness of the faint accretion disc plus gas stream emission (Fig.~\ref{fig_mapeamento}, lower panel). All reconstructions reached a unity final reduced chi-square.

The statistical uncertainties in the eclipse maps are estimated with a Monte Carlo procedure \citep[see e.g.][]{Rutten1992}. For a given input data curve, a set of $10^2$ artificial curves is generated in which the data points are independently and randomly varied according to a Gaussian distribution with standard deviation equal to the uncertainty at that point. The artificial curves are fitted with the eclipse-mapping algorithm to produce a set of randomized eclipse maps. These are combined to produce an average map and a map of the residuals with respect to the average, which yields the statistical uncertainty at each pixel. The uncertainties obtained with this procedure are used to plot the confidence level contours in the eclipse maps of Fig.~\ref{fig_mapeamento} and to estimate the errors in the derived radial temperature distributions.

Data and model light curves and corresponding eclipse maps in a logarithmic grayscale are shown in Fig.~\ref{fig_mapeamento}. The upper panel displays the eclipse map of the full light curve. The brightness distribution is dominated by the contribution from the WD at disc center. Aside of the deep, narrow eclipse of the WD, the additional broad, shallow and asymmetric eclipse shape maps into two asymmetric brightness sources, one running along the disc rim at $R_d \simeq 0.65\,R_{L_1}$ (signalling the presence of a weak bright spot), and the other extending along the ballistic stream trajectory beyond the impact point at disc rim (suggesting the presence of gas stream overflow/penetration). The fact that this residual bright spot leads to no perceptible orbital hump in the light curve suggests that at least the outer disc regions are optically thin. The lower panel shows the WD-subtracted light curve and corresponding eclipse map, revealing a faint disc at the center of the map as well as the two above mentioned asymmetric sources. These three brightness sources are statistically significant at the $2.5\sigma$ (bright spot) and 3-5\,$\sigma$ (disc plus gas stream emission) confidence levels. We find small and non significant uneclipsed components of $0.002 \pm 0.001$\,mJy for both the full and the WD-subtracted light curves.

\begin{figure}
	\includegraphics[width=\columnwidth]{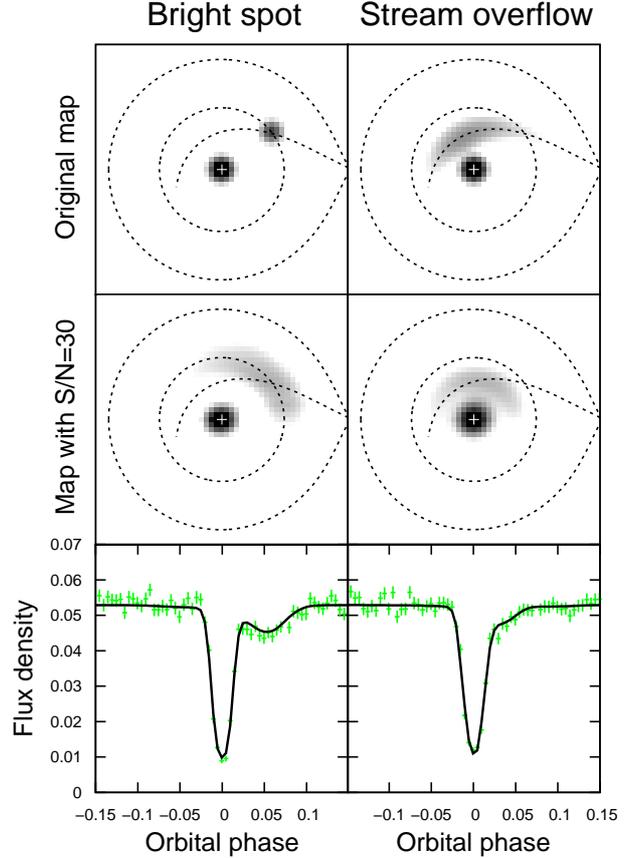}
    \caption{Reconstructions of asymmetric brighness sources with light curves of $S/N=30$. The uppermost panel shows the grayscale maps of the models with a bright spot (left) and with extended emission along the ballistic stream trajectory (right). The notation is the same as that of Fig.~\ref{fig_mapeamento}.}
    \label{fig_comparação}
\end{figure}

We performed simulations in order to gauge the ability of our eclipse mapping code to distinguish between a bright spot at disc rim and extended emission along the ballistic stream trajectory. For this purpose, we created two artificial brightness distributions, one with a WD at disc center plus a bright spot at the intersection of the ballistic stream trajectory with a disc of radius $R_d= 0.485\, R_{L1}$ (C11), and the other with enhanced emission along the ballistic stream trajectory well beyond the outer disc rim. Grayscale maps of these distributions are shown in the uppermost panels of Fig.~\ref{fig_comparação}. They were convolved with the eclipse geometry of J0926 to generate synthetic light curves to which Gaussian noise was added in order to simulate real data. Each of these light curves was then subjected to the eclipse mapping code. The lower and middle panels of Fig.~\ref{fig_comparação} show, respectively, the synthetic data and model light curves, and corresponding eclipse maps for the case of $S/N=30$, comparable to that of our light curves.

Because of the small mass ratio of J0926, the projected shadow of its tiny mass-donor star in the orbital plane is so narrow that it only starts covering the bright spot when the WD has almost fully recovered from its eclipse \citep[see Fig.\,3 of][]{Marsh2007}, thereby leading to two almost separate eclipses (Fig.~\ref{fig_comparação}, left-hand panels). This eclipse shape is reminiscent of the light curves of C11 and of the first two nights of observations of Z14, which show a significant orbital hump intrepreted in terms of anisotropic emission from a bright spot `painted' at disc rim. On the other hand, the extended emission of the `stream overflow' distribution is eclipsed together with the WD at disc center. Because of its extended distribution and asymmetric location, its eclipse lasts longer and is displaced towards positive phases with respect to that of the WD (Fig.~\ref{fig_comparação}, right-hand panels). The two artificial brightness distributions lead to different eclipse shapes, which are easy to distinguish even at relatively low S/N. As a consequence of the azimuthal blur of the maximum entropy eclipse mapping method \citep[see][]{Baptista2016}, the `bright spot' is reconstructed as an arc extended by $\simeq 90^\circ \, (= 3\,\Delta\theta)$ in azimuth at the correct radius, while the `stream overflow' is smeared in azimuth along the ballistic stream trajectory. These simulations show that the data allow us to easily distinguish between a compact bright spot at disc rim and an extended emission along the stream trajectory; they give us confidence to argue that there is evidence of gas stream overflow/penetration in J0926 at the epoch of our observations.

We assumed a distance of ($465\pm 5$)\,pc (C11) to convert the intensities of the eclipse maps to blackbody brightness temperatures, $T_b$, which may then be compared to the $T_\mathrm{eff}(R) \propto R^{-3/4}$ law of opaque, steady-state disc models, as well as to allow a rough estimate of the disc mass input rate $\dot{M}$. It is worth noting that $T_b$ may not be a proper estimate of the disc \& stream gas effective temperature, and that a relation between these quantities can only be obtained by self-consistent modelling of the atmosphere of these sources \citep{Baptista1998}. Additionally, interstellar extinction effects were not taken into account in this procedure, as the insterstellar extinction towards J0926 is expected to be negligible (C11) due to its high galactic latitude ($+46^\circ$). We separated the emission of the gas stream from that of the remaining disc by dividing the eclipse map into two azimuthal `slices of pizza' \citep[e.g.,][]{Baptista2000}, one for $\theta=0^\circ-90^\circ$ \footnote{$\theta$ is the azimuth measured with respect to the line joining both stars and increases counter-clockwise in the eclipse maps of Fig.~\ref{fig_mapeamento}.} (gas stream) and another for the remaining azimuths ($\theta=90^\circ-360^\circ$), and by computing median brightness temperatures in anulli of width $0.05\,R_{L_1}$ for each slice.

\begin{figure}
	\includegraphics[width=\columnwidth]{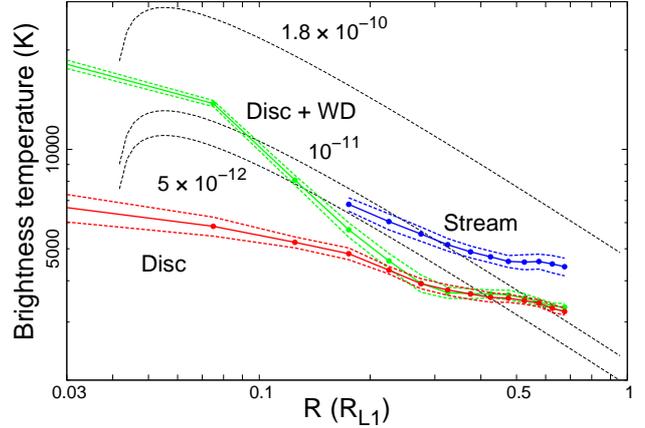}
    \caption{Azimuthally averaged radial brightness temperature distributions for the gas stream ($\theta=0^\circ-90^\circ$, blue), and for the full (`disc+WD', green) and WD-subtracted (`disc', red) eclipse maps ($\theta= 90^\circ-360^\circ$) for an assumed distance of 465\,pc. The dotted curves show the 1$\sigma$ limits on the corresponding distributions. Effective temperature distributions of opaque, steady-state disc models for mass accretion rates of $5\times 10^{-12}, 10^{-11}$, and $1.8\times 10^{-10}\,M_\odot\,yr^{-1}$ are shown as dotted lines.}
    \label{fig_radial}
\end{figure}

Azimutally averaged radial brightness temperature distributions of the combined (disc+stream+WD) and of the WD-subtracted surface brightness distributions are shown in Fig.~\ref{fig_radial}. The WD at disc center has $T_b = (19\pm 1)\times 10^3\,K$, consistent at the 2-$\sigma$ limit with the assumed WD temperature and indicating that $T_b \simeq T_\mathrm{eff}$ in this case. It is surrounded by a faint, cool accretion disc ($T_b \simeq 6000$-$3500\,K$ in the range $0.1$-$0.5\,R_{L_1}$) with enhanced emission along a hotter gas stream ($T_b\simeq 7000$-$4500\,K$).

The observed disc radial temperature distribution is much flatter than the $T\propto R^{-3/4}$ law and is reminiscent to those of quiescent dwarf novae \citep[e.g., OY~Car,][]{Wood1989}. From the brightness temperatures in the outer disc regions ($R = 0.5$-$0.6\,R_{L1}$) we estimate a disc mass input rate of $\dot{M} =(9 \pm 1)\times 10^{-12}\,M_\odot \,yr^{-1}$, more than an order of magnitude lower than the $\dot{M}$ expected from binary evolution with conservative mass transfer ($\simeq 1.8\times 10^{-10}\,M_\odot \,yr^{-1}$, Z14). We put forward a few possible explanations for this discrepancy. It may be that the outer disc is optically thin and the gas effective temperatures (and corresponding $\dot{M}$ values) are significantly larger than the inferred $T_b$ values; multicolour (or spectral) eclipse mapping would be helpful to clarify this issue. It may also be that the observed period increase is not a consequence of evolution with conservative mass transfer, but part of a cyclical period change which could hide a long term period increase at a much slower pace (a $\dot{P} \sim 1.5\times 10^{-14} \, s/s$ would be expected from the above mass transfer rate estimate). A third explanation would be that the present mass transfer rate of J0926 is significantly lower than its secular average. This raises the possibility of the existence of large amplitude, long-term modulations in mass transfer rates of AM CVn stars similar to those which presumably occur in cataclysmic variables \citep[the hibernation scenario of novae, see][]{Kovetz1988}. However, while in CVs these modulations seem a consequence of their recurrent nova eruptions, in AM CVn stars a different mechanism would be required since their hydrogen-deficient discs cannot drive hydrogen-burning thermonuclear runaways at the WD surface.

\section{Conclusions}\label{summary}

We observed J9026 at five occasions on 2012 February-March, during a 4.6\,yr long period without recorded outbursts. The light curves of all nights consistently show the same morphology, brightness level, eclipse shape and depth, with no evidence of superhumps or of an orbital hump produced by anisotropic emission from a bright spot at disc rim. We combined our median eclipse timing with those in the literature to revise the binary ephemeris and to confirm that the orbital period is increasing at a rate $\dot{P}=(3.2\pm 0.4)\times 10^{-13} \, s/s$. Eclipse mapping of the average light curve shows a hot white dwarf at disc centre surrounded by a faint, cool accretion disc with a residual bright spot (at $R_d\simeq 0.65\,R_{L1}$) and enhanced emission along the ballistic stream trajectory well beyond  the disc rim, suggesting the occurence of gas stream overflow or penetration at that epoch. For an assumed distance of 465\,pc, the estimated disc mass input rate of $\dot{M}=(9 \pm 1)\times 10^{-12}\, M_\odot\,yr^{-1}$ is more than an order of magnitude lower than that expected from binary evolution with conservative mass transfer.

\section*{Acknowledgements}

We are grateful to Michael Bode for granting LRT time for these observations, and to Iain Steele and Robert Smith for much appreciated help with the data reduction process. W. S. acknowledges financial support from INCT-A/Brazil and CNPq/Brazil. This paper is based on observations made with the Liverpool Telescope operated on the island of La Palma by Liverpool John Moores University in the Spanish Observatorio del Roque de los Muchachos of the Instituto de Astrofisica de Canarias with financial support from the UK Science and Technology Facilities Council. This research has made use of the SIMBAD database, operated at CDS, Strasbourg, France.







\bsp	
\label{lastpage}
\end{document}